\documentstyle[aps,pra,epsfig,multicol]{revtex}

\begin{document}

\draft

\title{Quantum Characterization of a Werner-like Mixture}

\author{Anna Felicetti, Stefano Mancini and
Paolo Tombesi}

\address{INFM, Dipartimento di Fisica, Universit\`a di Camerino,
I-62032 Camerino, Italy
}

\date{\today}

\maketitle

\begin{abstract}
We introduce a Werner-like mixture 
[R. F. Werner, Phys. Rev. A {\bf 40}, 4277 (1989)] 
by considering two correlated but different degrees of freedom,
one with discrete variables and the other 
with continuous variables.
We evaluate the mixedness 
of this state, and its degree of entanglement
establishing its usefulness for quantum information 
processing like quantum teleportation. 
Then, we provide its
tomographic characterization.  
Finally, we show how such a mixture can be 
generated and measured in a trapped system like
one electron in a Penning trap.
\end{abstract}

\pacs{Pacs No: 03.65.Wj, 03.65.Ud, 03.65.Ta}

\begin{multicols}{2}

\section{Introduction}

It is nowadays well known that the nonlocal properties of Quantum 
Mechanics \cite{epr,bell} 
enable striking processes in quantum information \cite{qibook}.
In these processes it is prominent the 
role of maximally entangled states \cite{peres,vedral}.
However, very often, the decoherence effects due to the environment 
transform the pure entangled state into a statistical mixture and 
degrade 
quantum entanglement in the real world \cite{zur}.
Although purification schemes may be applied to noisy channels 
\cite{purif},
there exist some mixture states which maintain interesting properties.
An illuminating example is provided by
the Werner mixture \cite{werner}, which is not a mixture of 
product states, nonetheless not violating any Bell's inequality 
\cite{bell}, 
but still useful for quantum information processing 
\cite{popescu,benmix}.  
Such states belong to systems with 
two discrete degrees of freedom
like two spin-$\frac{1}{2}$.
However, information processing may sometimes involve 
hybrid systems where one degree of freedom has discrete 
variables and the other continuous variables.
It may happen, for instance, in trapped ions \cite{nist,win1},
or in trapped electrons \cite{ana}
or in cavity quantum electrodynamics \cite{cqed,haroche}.
Thus, it will be the aim of this paper to consider a mixture, which 
resembles the Werner one, but one of the two 
subsystems is described by continuous variables. 

On the other hand, states and processes used in quantum information 
typically need of a well characterization \cite{qibook}. 
This can be accomplished by using tomographic techniques \cite{vogris}.
Concerning the quantum state measurement, 
after the seminal work by Vogel and Risken \cite{vogris},
a lot of progress has been obtained and
further techniques and algorithms 
were developed \cite{tomrev}.
We would just mention the possibility of 
state reconstruction, for a composite system
of discrete and continuous variables,
by simply measuring the set of 
rotated spin projections and displaced number operators,
\cite{michol,njp}.  
Then, we shall provide 
the tomographic characterization of a Werner-like mixture
by generalizing that method.

The outline of the paper is the following:
In Section II we discuss the Werner mixture
and we extend the concept by considering one of the two 
subsystems with continuous variable.
Then we characterize such a state in terms of 
mixedness and entanglement.
Section III is devoted to the tomographic method employed 
for such a state reconstruction.
In Section IV we present the results of numerical simulations.
Finally, in Section V we discuss a possible implementation and 
Section VI is devoted to conclusions.

\section{Werner-like mixture}

In his pioneering paper, Bell proved that a local realistic 
interpretation of Quantum Mechanics is impossible \cite{bell},
and for the case of pure states  it is known that, when 
measurements are performed on two quantum systems separated 
in space, their results are correlated in a manner which, 
in general, cannot be explained by a local hidden variables
model \cite{gis}. Since the only pure states satisfying the Bell
inequality are pure product states,  one might naively
think that the only mixed states that do not violate Bell's 
inequality are mixtures of product states. However, Werner 
\cite{werner} showed that this conjecture is false for the 
so-called Werner states 
\begin{equation}\label{eq:rhower}
\rho=\frac{1}{8} I_{1} \otimes I_{2}
+\frac{1}{2}|\Psi^-\rangle\langle\Psi^-|\,,
\end{equation}
where $I_{i}$ ($i=1,2$) stands for the identity operator
of a single qubit 
($|\downarrow\rangle_{i}{}_{i}\langle\downarrow|
+|\uparrow\rangle_{i}{}_{i}\langle\uparrow|$) and 
$|\Psi^-\rangle=\frac{1}{\sqrt{2}}\left(
|\downarrow\rangle_{1}|\uparrow\rangle_{2} 
-|\uparrow\rangle_{1}|\downarrow\rangle_{2}\right)$
is the spin singlet state.

A more general Werner mixture can be obtained by
considering one of the two subsystems,
say $2$, as described by continuous variables.
A way to encode qubit in continuous variable systems
could be the use of even and odd cat states which are
orthogonal \cite{coc}, thus resulting in the same situation 
of Eq.(\ref{eq:rhower}).
Instead, the choice we are going to make
is more general and gives the possibility 
to explore a variety of situations.

\begin{figure}[t]
\centerline{\epsfig{figure=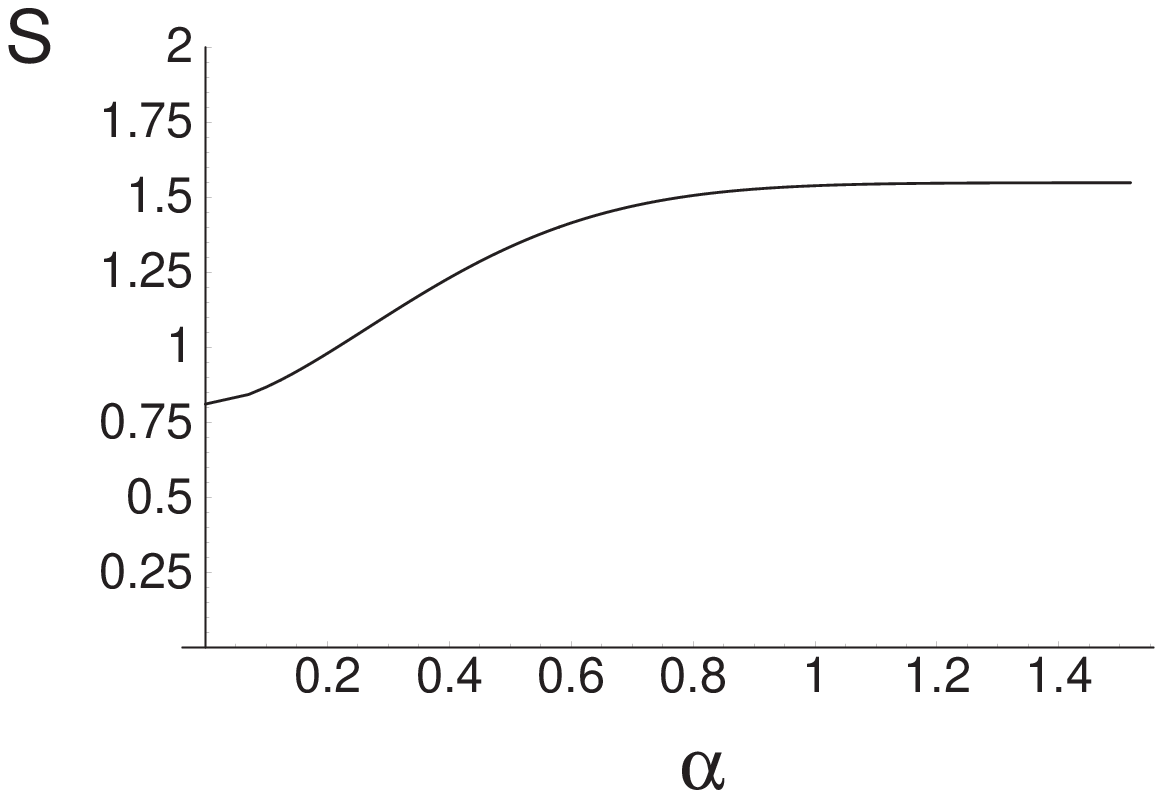,width=3.0in}}
\caption{\narrowtext 
Entropy $S$ as function of $\alpha\equiv\sqrt{-(1/2)\ln\kappa}$.
}
\label{fig1}
\end{figure}

That is, we now replace the states $|\uparrow\rangle_{2}$, 
$|\downarrow\rangle_{2}$ of the second qubit with
$|\alpha\rangle_{2}$ and $|-\alpha\rangle_{2}$, where 
the latter are coherent states of amplitude $\alpha$
and $-\alpha$ respectively 
(we shall consider $\alpha\in{\bf R}$ throughout the paper for the
sake of simplicity).
Therefore, a Werner-like mixture would be
\begin{eqnarray}\label{eq:rhoal}
\rho&=&\frac{1}{8}\left\{
|\downarrow\rangle_{1}{}_{1}\langle\downarrow|
\otimes |-\alpha\rangle_{2}{}_{2}\langle-\alpha|\right.
\nonumber\\
&&\left.
+|\uparrow\rangle_{1}{}_{1}\langle\uparrow|
\otimes |-\alpha\rangle_{2}{}_{2}\langle-\alpha|\right.
\nonumber\\
&&\left.
+|\downarrow\rangle_{1}{}_{1}\langle\downarrow|
\otimes |\alpha\rangle_{2}{}_{2}\langle\alpha|\right.
\nonumber\\
&&\left.
+|\uparrow\rangle_{1}{}_{1}\langle\uparrow|
\otimes |\alpha\rangle_{2}{}_{2}\langle\alpha|\right\}
\nonumber\\
&&+\frac{1}{4}\left[|\downarrow\rangle_{1}|\alpha\rangle_{2}
-|\uparrow\rangle_{1}|-\alpha\rangle_{2}\right]
\left[{}_{1}\langle\downarrow|{}_{2}\langle\alpha|
-{}_{1}\langle\uparrow|{}_{2}\langle-\alpha|\right]\,.
\nonumber\\
\end{eqnarray}
Since $\langle\alpha|-\alpha\rangle=\exp(-2\alpha^{2})\ne 0$,
the above state doesn't describe a real two qubit system,
but rather a two qubit system with nonorthogonal states \cite{hirota}.
Of course, for $\alpha \gg 1$ Eq.(\ref{eq:rhoal}) behaves like the 
state
(\ref{eq:rhower}), but we want to study its characteristics 
for a generic value of $\alpha$.
To this end, we map the above state in the two spin-$\frac{1}{2}$
Hilbert spaces by introducing for the subsystem $2$ a vector 
$|\psi\rangle_{2}=\kappa|\downarrow\rangle_{2}$
$+\sqrt{1-\kappa^{2}}|\uparrow\rangle_{2}$ and considering the
non orthogonal states
$|\downarrow\rangle_{2}\,,|\psi\rangle_{2}$ 
instead of $|-\alpha\rangle_{2}\,,|\alpha\rangle_{2}$, with 
$\langle\psi|\downarrow\rangle=\kappa\equiv\exp(-2\alpha^{2})$.
Then, Eq.(\ref{eq:rhoal}) can be rewritten as
\begin{eqnarray}\label{eq:rhopsi}
\rho&=&\frac{1}{8}\left\{
|\downarrow\rangle_{1}{}_{1}\langle\downarrow|
\otimes |\downarrow\rangle_{2}{}_{2}\langle\downarrow|\right.
\nonumber\\
&&\left.
+|\uparrow\rangle_{1}{}_{1}\langle\uparrow|
\otimes |\downarrow\rangle_{2}{}_{2}\langle\downarrow|\right.
\nonumber\\
&&\left.
+|\downarrow\rangle_{1}{}_{1}\langle\downarrow|
\otimes |\psi\rangle_{2}{}_{2}\langle\psi|\right.
\nonumber\\
&&\left.
+|\uparrow\rangle_{1}{}_{1}\langle\uparrow|
\otimes |\psi\rangle_{2}{}_{2}\langle\psi|\right\}
\nonumber\\
&&+\frac{1}{4}\left[|\downarrow\rangle_{1}|\psi\rangle_{2}
-|\uparrow\rangle_{1}|\downarrow\rangle_{2}\right]
\left[{}_{1}\langle\downarrow|{}_{2}\langle\psi|
-{}_{1}\langle\uparrow|{}_{2}\langle\downarrow|\right]
\,.
\nonumber\\
\end{eqnarray}
The degree of mixedness of the state (\ref{eq:rhopsi}) 
can be evaluated by using
the Von Neumann entropy \cite{von}
\begin{equation}\label{eq:Sdef}
S(\rho)=-{\rm Tr}\left(\rho \log_{2} \rho\right)
=-\sum_{i}\lambda_{i}\log_{2}\lambda_{i}\,,
\end{equation}
where $\lambda_{i}$ are the eigenvalues of the matrix representation 
of $\rho$. 
We have calculated such eigenvalues in the basis 
${\cal B}\equiv$
$\{|\downarrow\rangle_{1}|\downarrow\rangle_{2}$,
$|\downarrow\rangle_{1}|\uparrow\rangle_{2}$,
$|\uparrow\rangle_{1}|\downarrow\rangle_{2}$,
$|\uparrow\rangle_{1}|\uparrow\rangle_{2}\}$
and we have plotted the entropy $S$ in Fig.\ref{fig1}.

We can see that also for $\alpha=0$ the state is a mixture,
then, it becomes more and more mixed by increasing 
the value of $\alpha$, but never reaching the maximum
(a completely mixed density operator in a $d$-dimensional 
space has entropy $\log_{2} d$).
It is also worth noting that for $\alpha$ ($\kappa$) arbitrary, 
Eq.(\ref{eq:rhopsi}) is not
a mixture of Bell's states 
as it is for the Werner mixture (\ref{eq:rhower}), i.e., 
for $\alpha\gg 1$. 

\begin{figure}[t]
\centerline{\epsfig{figure=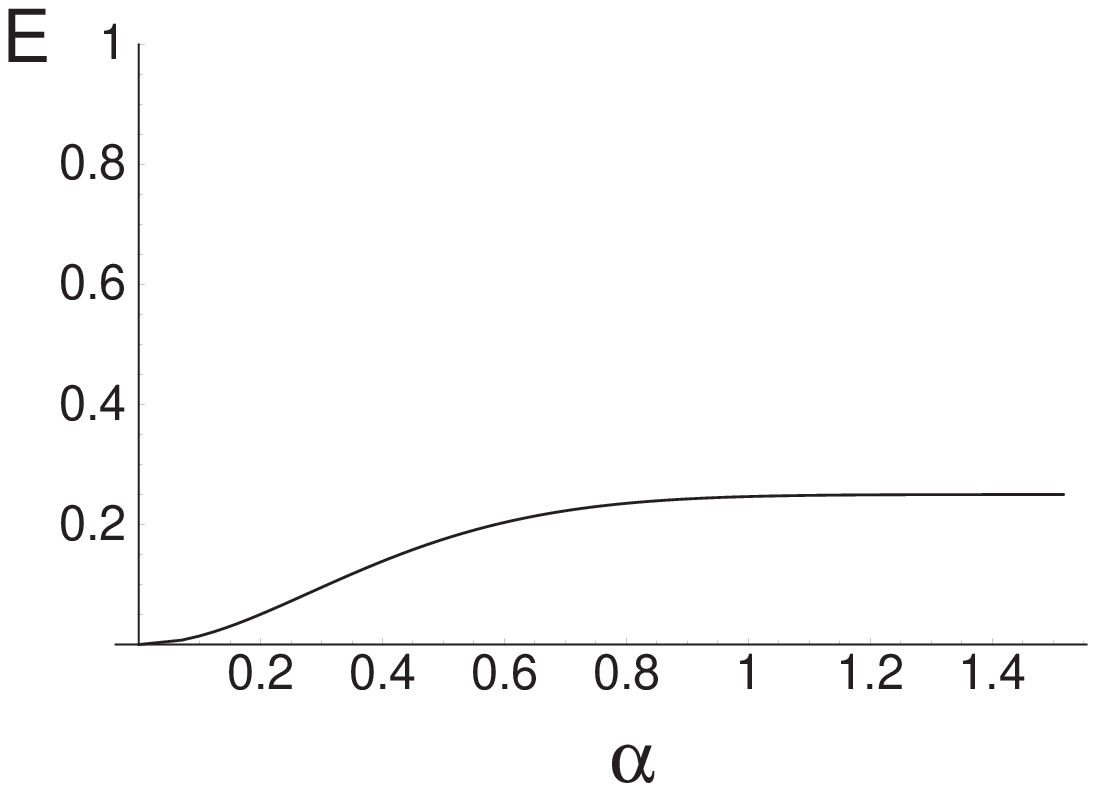,width=3.0in}}
\caption{\narrowtext 
Degree of entanglement $E$ 
as function of $\alpha\equiv\sqrt{-(1/2)\ln\kappa}$.
}
\label{fig2}
\end{figure}

Let us now consider the measure of entanglement.
For a two spin-$\frac{1}{2}$ system the state separability 
is related to the partial transposition operation \cite{peres,perhor}.
The matrix elements of partial transposition 
$\rho^{T_{2}}$ of a state $\rho$ are given by
$\rho_{m\mu,\,n\nu}^{T_{2}}\equiv\rho_{m\nu,\,n\mu}$
where $\rho_{m\mu,\,n\nu}={}_{1}\langle m|{}_{2}\langle \mu|
\rho |\nu\rangle_{2} |n\rangle_{1}$
with $m,n,\mu,\nu=\uparrow,\downarrow$.
A density matrix $\rho$ for a two spin-$\frac{1}{2}$ system is 
inseparable if and only if its partial transpose,
$\rho^{T_{2}}$, 
has any negative eigenvalue \cite{peres,perhor}.
Then, a suitable measure of entanglement can be defined as \cite{leekim}
\begin{equation}\label{eq:Edef}
E(\rho)=-2\sum_{i}\lambda_{i}^{-}\,,
\end{equation}
where $\lambda_{i}^{-}$ is a negative eigenvalue of $\rho^{T_{2}}$.
It is worth noting that this measure satisfies the necessary conditions
required for every measure of entanglement \cite{vedral}.
Then, we have plotted in Fig.\ref{fig2} 
the quantity $E$ calculated by exploiting
again the matrix representation of Eq.(\ref{eq:rhopsi}) in the 
basis ${\cal B}$.

\begin{figure}[t]
\centerline{\epsfig{figure=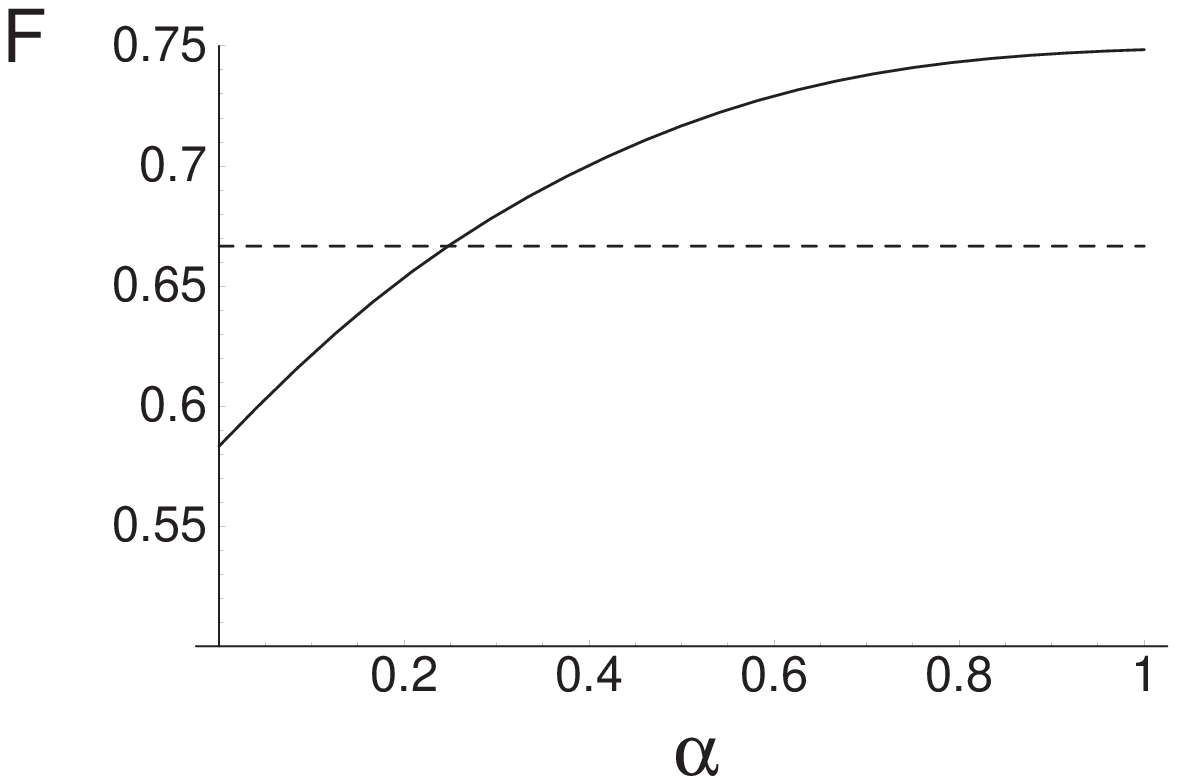,width=3.0in}}
\caption{\narrowtext 
Teleportation fidelity as function of 
$\alpha\equiv\sqrt{-(1/2)\ln\kappa}$.
The dashed line represents the classical bound of $2/3$.
}
\label{fig3}
\end{figure}

We can see that the Werner-like mixture is factorisable for
$\alpha=0$, then, by increasing $\alpha$, 
its degree of entanglement increases saturating
at the value $1/4$ characteristic of the true Werner mixture 
(\ref{eq:rhower}).
This latter value of entanglement
is known to be sufficient 
to improve the teleportation performances
over classical limit
once the sender (Alice) and the receiver (Bob)
initially share the state (\ref{eq:rhower})
\cite{popescu}. Thus, it is straightforward 
to ask to what extent a Werner-like mixture 
can be used for the same goal.
To establish a threshold value for $\alpha$
we are going to consider the teleportation fidelity.
To this end, we first write
the state $\rho$ through the Hilbert-Schmidt 
decomposition
\begin{eqnarray}\label{eq:rhoHS}
\rho=\frac{1}{4}&&\left[
I_{1} \otimes I_{2}
+{\bf r}{\bf \cdot} \mbox{\boldmath $\sigma$}^{(1)} \otimes I_{2}
+I_{1} \otimes {\bf s}{\bf \cdot}\mbox{\boldmath $\sigma$}^{(2)}\right.
\nonumber\\ 
&&\left.
+\sum_{n,m=1}^{3} t_{n,m} \sigma_{n}^{(1)} \otimes \sigma_{m}^{(2)}
\right]\,,
\end{eqnarray}
where $\{\sigma_{n}\}_{n=1,2,3}$ are the standard Pauli matrices,
${\bf r}$, ${\bf s}$ are vectors in ${\bf R}^{3}$
and ${\bf r}{\bf\cdot}\mbox{\boldmath $\sigma$}
=\sum_{i=1}^{3}r_{i}\sigma_{i}$.
Furthermore, the coefficients $t_{n,m}={\rm Tr}[\rho\,\sigma_{n}^{(1)}
\otimes \sigma_{m}^{(2)}]$ form the real matrix $T$ describing the 
correlations between the two qubits.   
Thus, the teleportation capabilities will depend on the specific 
form of $T$. In particular it is shown in Ref.\cite{horo} that the 
teleportation fidelity amounts to 
\begin{equation}\label{eq:Fdef}
F=\frac{1}{2}\left[1+\frac{1}{3}{\rm Tr}\sqrt{T^{\dag}T}\right]\,.
\end{equation}
Then, in Fig.\ref{fig3} we have plotted the quantity $F$ versus 
$\alpha$ compared with the classical fidelity $2/3$.

We immediately recognize the presence of a threshold value 
($\alpha=0.2476$) below which the Werner-like mixture 
becomes useless for quantum teleportation.

\section{State measurement}

We now discuss the possibility of a complete characterization
of the Werner-like mixture through tomographic techniques.
In particular, we generalize the method presented in 
Refs.\cite{michol,njp} to non pure states.

Accordingly to the state reconstruction principle developed in 
Ref.\cite{jmo} we choose an observable, here
$\sigma_{3}\otimes a^{\dag}a$, then we apply
suitable unitary transformations to get a set of observables
giving the whole state information upon measurements.
In our case the transformations would be
\begin{eqnarray}
U(\theta,\varphi)&=&\exp\left[-i\theta\left(
\sigma_{1}\cos\varphi+\sigma_{2}\sin\varphi\right)\right]\,,
\label{eq:U}\\
D(\beta)&=&\exp\left[\beta a^{\dag}-\beta^{*} a\right]\,,
\label{eq:D}
\end{eqnarray}
which lead to rotated (by angles $\theta$ and $\varphi$)
spin projection in the subsystem $1$ \cite{weig},
and to displaced number state (by a complex amount $\beta$) 
in the subsystem $2$ \cite{pnt}.
Then, it is possible to consider the following 
measurable marginal distributions
\begin{eqnarray}\label{eq:wdef}
w(s,n;\theta,\varphi,\beta)&=&
{\rm Tr}\left\{
\rho\,D(\beta)U(\theta,\varphi)\right.
\nonumber\\
&&\left.
|s\rangle_{1}{}_{1}\langle s|
\otimes |n\rangle_{2}{}_{2}\langle n|U^{\dag}(\theta,\varphi)
D^{\dag}(\beta)\right\}\,,
\nonumber\\
\end{eqnarray}
having as variables 
the eigenvalues $s$, $n$ of $\sigma_{3}$ and $a^{\dag}a$
as variables and parametrically
depending on $\theta$, 
$\varphi$ and $\beta$.
Thus, measuring the state $\rho$ would mean the possibility
to express $\rho$ as a functional operator of $w$,
i.e. to invert expression (\ref{eq:wdef}). 
This also means the possibility to 
sample the density
matrix elements (in some basis)
from the quantity measured by
spanning the whole space of parameters. 

In reality, we shall see that it is not necessary 
to consider all possible values of parameters.
As matter of fact, we write the density operator 
(\ref{eq:rhoal}) as
\begin{equation}\label{eq:rhomat}
\rho=\left[
\begin{array}{cc}
\rho^{\uparrow\uparrow}&\rho^{\uparrow\downarrow}
\\
\rho^{\downarrow\uparrow}&\rho^{\downarrow\downarrow}
\end{array}
\right]\,,
\quad
\rho^{\downarrow\uparrow}=[\rho^{\uparrow\downarrow}]^{\dag}\,,
\end{equation}
where each operator  
$\rho^{\uparrow\uparrow}$,
$\rho^{\uparrow\downarrow}$,
$\rho^{\downarrow\uparrow}$,
$\rho^{\downarrow\downarrow}$,
can in turn be represented in the Fock basis 
of the subsystem $2$.

Now, we set $\theta=\varphi=0$
and we suppose to retain only the 
measurement results $s=\uparrow$; then, 
expanding the density operator $\rho^{\uparrow\uparrow}$ 
in the Fock basis, 
and defining $N_{c}$ as an appropriate estimate of the maximum number 
of excitations (cut-off), we have
\begin{equation}\label{eq:wup}
w(\uparrow,n;0,0,\beta)=
\sum_{k,m=0}^{N_{c}} \langle n, \beta
|k\rangle\langle k|\rho^{\uparrow\uparrow}|m\rangle\langle 
m|n,\beta\rangle\,.
\end{equation}
The projection of the displaced number state $|n,\beta\rangle$ 
onto the Fock state $|m\rangle$ can be obtained generalizing the 
result derived in Ref.~\cite{cah}.

Let us now consider, for a given value of $|\beta|$, $w$
as a function of $\phi=\arg[\beta]$ \cite{opa} 
and calculate the coefficients of the Fourier expansion
\begin{equation}\label{eq:FTwup}
w(\uparrow,n;0,0,r) = \frac{1}{2\pi}\int_{0}^{2\pi} d\phi \;
w(\uparrow,n;0,0,\phi) e^{ir\phi}\;,
\end{equation}
for $r=0,1,2,\ldots$. Combining Eqs.~(\ref{eq:wup}) and 
(\ref{eq:FTwup}), we get
\begin{equation}\label{eq:wGrho}
w(\uparrow,n;0,0,r) = \sum_{m=0}^{N_{c}-r} {G}^{(r)}_{n,m}
(|\beta|) \langle m+r | \rho^{\uparrow\uparrow} | m \rangle\,,
\end{equation}
where the explicit expression of the matrices $G$ is given in
Ref.~\cite{michol}.

We may now notice that if the 
distribution $w(\uparrow,n;0,0,\beta)$ is 
measured for $n\in [0,N]$ with $N\geq N_{c}$, then 
Eq.~(\ref{eq:wGrho}) represents, for each value of $r$, a system of 
$N+1$ linear equations between the $N+1$ measured quantities and the 
$N_{c}+1-r$ unknown density matrix elements. Therefore, in order to 
obtain the latter, we only need to invert the system
\begin{equation}\label{eq:rhoGw}
\langle m+r | \rho^{\uparrow\uparrow} | m \rangle = \sum_{n=0}^{N}
M^{(r)}_{m,n} (|\beta|) w(\uparrow,n;0,0,r)\,,
\end{equation}
where the matrices $M$ are given by $M=(G^{T} G)^{-1} G^{T}$.

The procedure can be repeated with $w(\downarrow,n;0,0,\beta)$
in order to get the matrix elements of $\rho^{\downarrow\downarrow}$.
Then, changing the parameters so that $\theta=\pi/4$ and 
$\varphi=-\pi/2$ we can get the real part of
the matrix elements of $\rho^{\uparrow\downarrow}$.
Instead, with $\theta=\pi/4$ and 
$\varphi=0$ we can get the imaginary part of
the matrix elements of $\rho^{\uparrow\downarrow}$, 
thus concluding the reconstruction procedure
of the state (\ref{eq:rhomat}).

\section{Numerical results}

\begin{figure}[t]
\centerline{\epsfig{figure=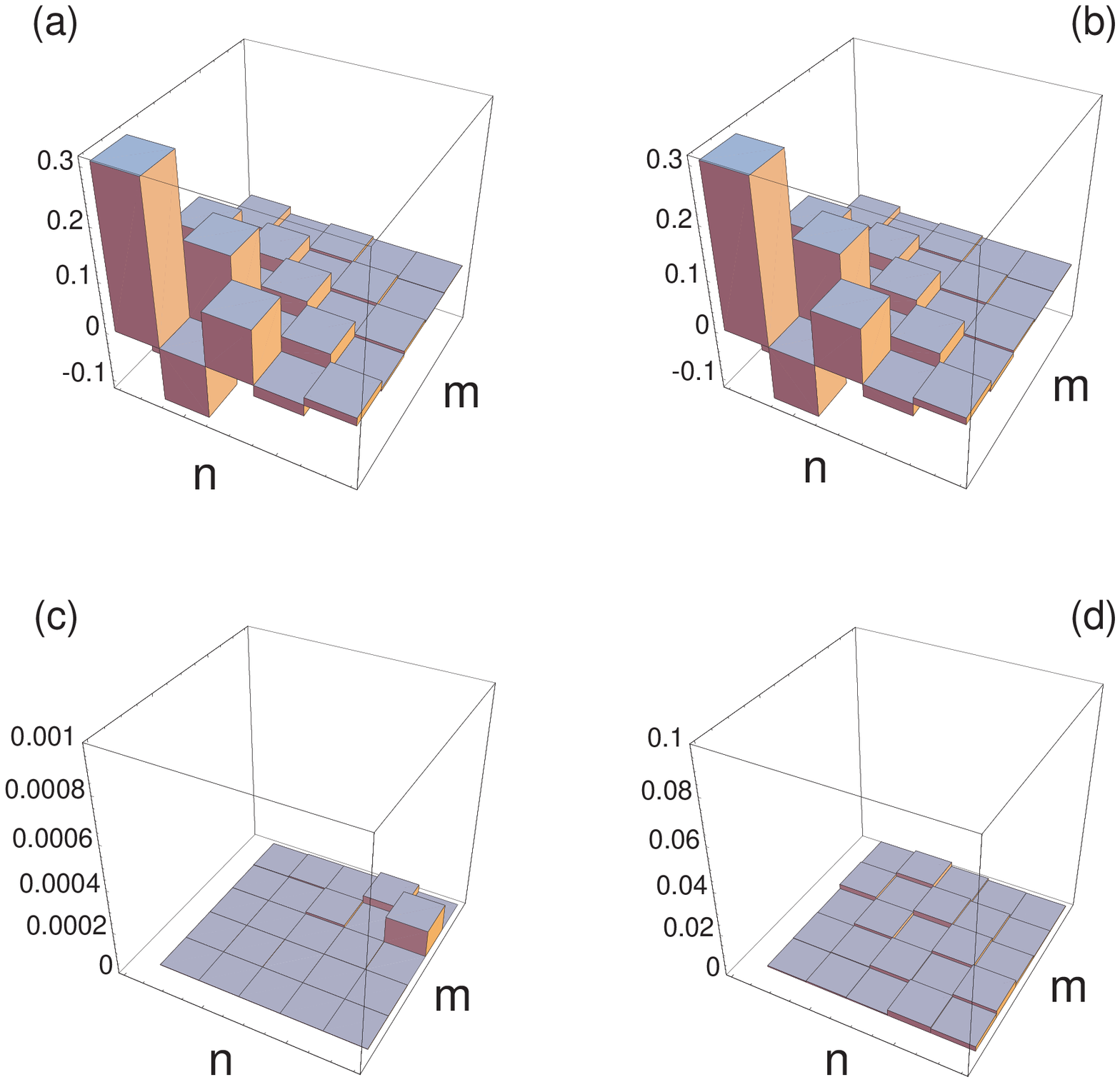,width=3.0in}}
\caption{\narrowtext 
Density matrix elements of $\rho^{\uparrow\uparrow}$
for the state (\ref{eq:rhoal}) with $\alpha=0.7$.
In the computer simulation
$N_{\phi}=96$ phases are considered and  
${\cal N}_{\phi}=10^{4}$ measurement events at each phase are 
assumed to be recorded using $N_{c}=31$.
The other parameters are $|\beta|=0.6$ and $\eta=0.9$.
(a) Exact density matrix;
(b) reconstructed density matrix;
(c) statistical errors;
(d) absolute difference between reconstructed and exact density
matrix elements.
}
\label{fig4}
\end{figure}

\begin{figure}[t]
\centerline{\epsfig{figure=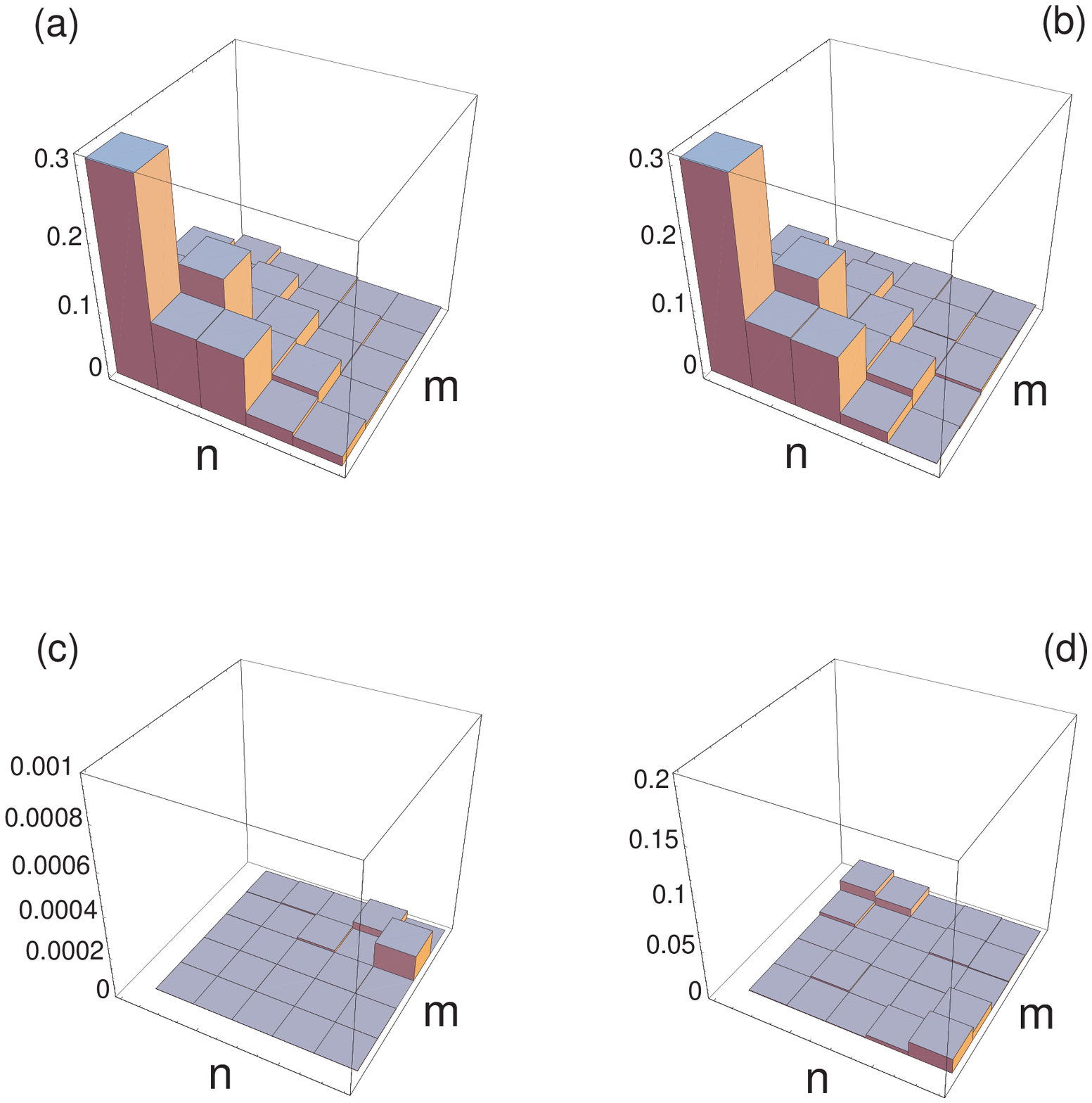,width=3.0in}}
\caption{\narrowtext 
Density matrix elements of $\rho^{\downarrow\downarrow}$
for the state (\ref{eq:rhoal}) with $\alpha=0.7$.
In the computer simulation
$N_{\phi}=96$ phases are considered and  
${\cal N}_{ev}=10^{4}$ measurement events at each phase are 
assumed to be recorded using $N_{c}=31$.
The other parameters are $|\beta|=0.6$ and $\eta=0.9$.
(a) Exact density matrix;
(b) reconstructed density matrix;
(c) statistical errors;
(d) absolute difference between reconstructed and exact density
matrix elements.
}
\label{fig5}
\end{figure}

\begin{figure}[t]
\centerline{\epsfig{figure=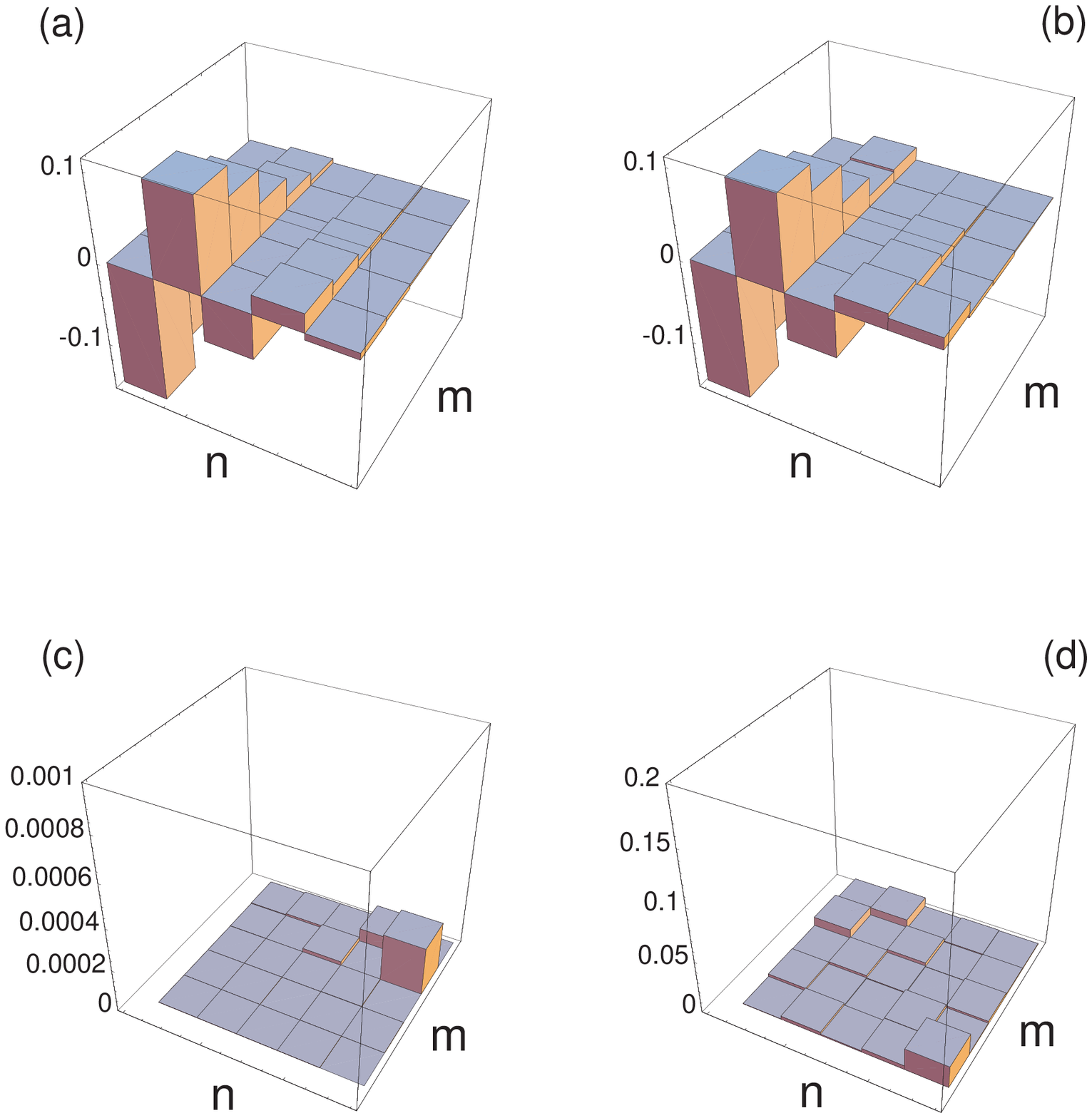,width=3.0in}}
\caption{\narrowtext 
Density matrix elements of $\rho^{\uparrow\downarrow}$
for the state (\ref{eq:rhoal}) with $\alpha=0.7$.
In the computer simulation
$N_{\phi}=96$ phases are considered and  
${\cal N}_{ev}=10^{4}$ measurement events at each phase are 
assumed to be recorded using $N_{c}=31$.
The other parameters are $|\beta|=0.6$ and $\eta=0.9$.
(a) Exact density matrix;
(b) reconstructed density matrix;
(c) statistical errors;
(d) absolute difference between reconstructed and exact density
matrix elements.
}
\label{fig6}
\end{figure}

As an example of the proposed method, 
we show in Figs.~\ref{fig4}, \ref{fig5}, \ref{fig6} 
the results of numerical Monte-Carlo 
simulations of the reconstruction of the
state~(\ref{eq:rhoal}) once written in the form (\ref{eq:rhomat}). 
In this simulation we have used the value $\alpha=0.7$
which make the state different from a true Werner mixture,
but still having nonclassical features
as discussed in Section II.

In order to account for experimental conditions, we 
have also considered the effects of a non-unit quantum efficiency 
$\eta$ in 
the counting of the number of excitations. When $\eta < 1$, the 
actually 
measured distribution is related to the ideal distribution by a 
binomial
convolution~\cite{scula}. 

Statistical errors are accounted for as well by considering an 
estimation of the marginal distributions given by 
$w^{\rm est}(n;\phi)={\cal N}_{\phi}(n)/{\cal N}_{\phi}$, where 
${\cal N}_{\phi}(n)$ is the number of events with $n$ counts at
phase $\phi$, while ${\cal N}_{\phi}$ is the total number of events 
at the same phase. Then,
following the arguments given in Refs.\cite{opa,leo}, the quantities
${\cal N}_{\phi}(n)$ can approximately be regarded as independent 
Poissonian random variables, whose means and variances are given by
$w(n;\phi){\cal N}_{\phi}\approx w^{\rm est}(n;\phi){\cal N}_{\phi}$.
The variance of $w^{\rm est}(n;\phi)$ may then be approximated by
$w^{\rm est}(n;\phi)/{\cal N}_{\phi}$, so that the variances of the 
real 
and imaginary parts of the density matrix can be easily estimated 
using Eq.(\ref{eq:rhoGw}) (and similarly for the other density 
operators).
This means that the errors can be estimated in real time as the 
experiment runs, simultaneously with the reconstruction of the density 
matrix elements.

Other error sources leading to discrepancies between true and 
reconstructed density matrices can be identified in the choice of
$|\beta|$, $N_{c}$ and $N_{\phi}$ (number of phases).
However, as it can be seen from 
Figs.~\ref{fig4}, \ref{fig5}, \ref{fig6} the
reconstructed density matrices turn out to be quite faithful. 

In addition to the present case
we have performed other simulations with different  
values of $\alpha$ 
and several values of the parameters, which confirm that the present 
method is quite stable and accurate. 

The phase-space description corresponding to~(\ref{eq:rhomat}) is given 
by the Wigner-function matrix~\cite{wig} 
\begin{eqnarray}\label{eq:wigmat}
W&=& \left[\begin{array}{cc}
W^{\uparrow\uparrow}(\gamma)
&
W^{\uparrow\downarrow}(\gamma)
\\
W^{\downarrow\uparrow}(\gamma)
&
W^{\downarrow\downarrow}(\gamma)
\end{array}
\right]
\nonumber\\
&=&\left[
\begin{array}{cc}
{\rm Tr}_{2}[\rho^{\uparrow\uparrow}\,\delta(\gamma -\hat{a})]
&
{\rm Tr}_{2}[\rho^{\uparrow\downarrow}\,\delta(\gamma -\hat{a})]
\\
{\rm Tr}_{2}[\rho^{\downarrow\uparrow}\,\delta(\gamma -\hat{a})]
&
{\rm Tr}_{2}[\rho^{\downarrow\downarrow}\,\delta(\gamma -\hat{a})]
\end{array}
\right]\,,
\end{eqnarray}
where $\delta(\gamma -\hat{a})$
is the Fourier transform of the displacement operator \cite{cah}.

\begin{figure}[t]
\centerline{\epsfig{figure=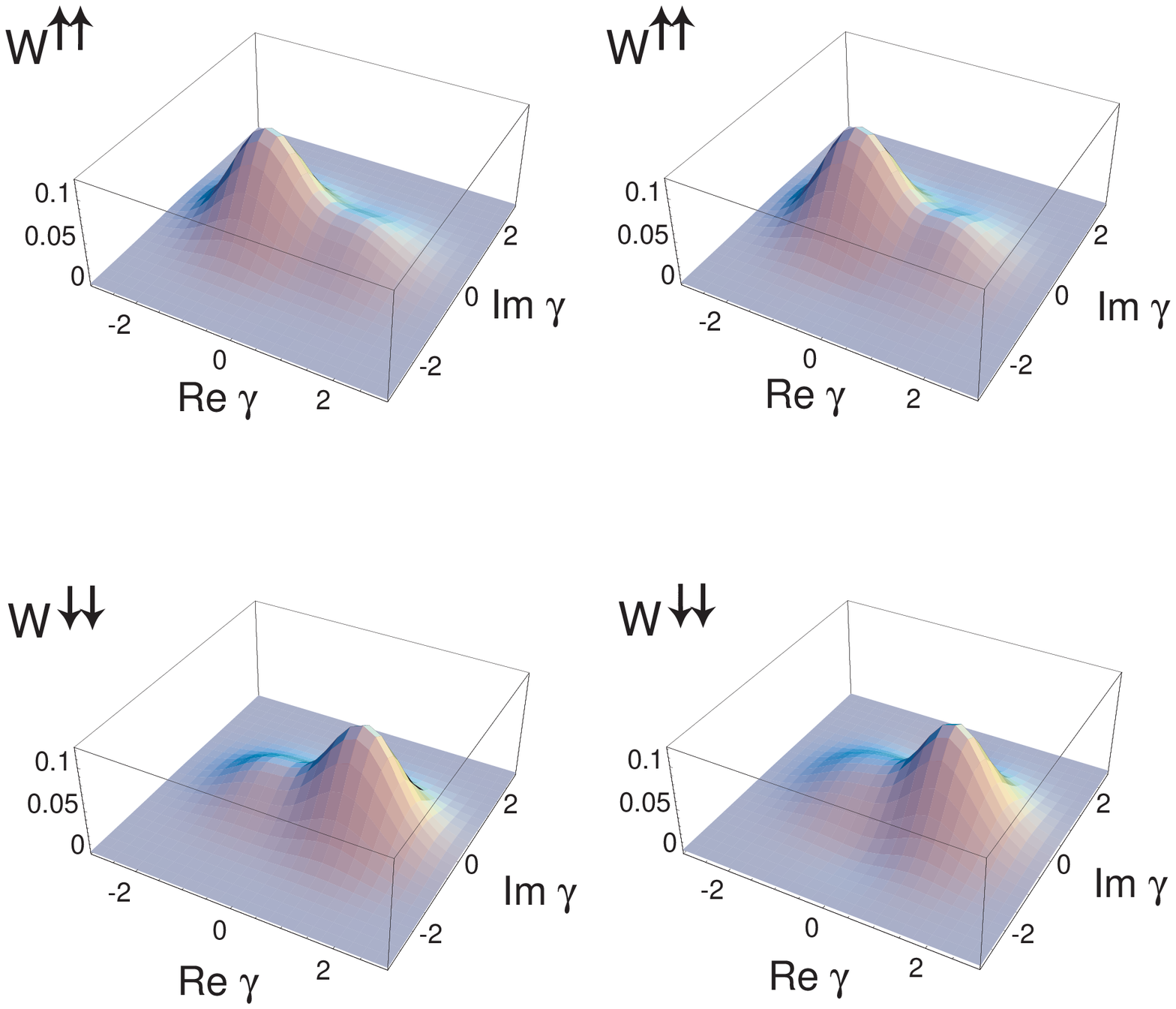,width=3.0in}}
\caption{\narrowtext
Top: Wigner function corresponding to the density matrix of 
Fig.\ref{fig4} (true on the left and reconstructed on the right).
Bottom: Wigner function corresponding to the density matrix of 
Fig.\ref{fig5} (true on the left and reconstructed on the right).
}
\label{fig7}
\end{figure}

\begin{figure}[t]
\centerline{\epsfig{figure=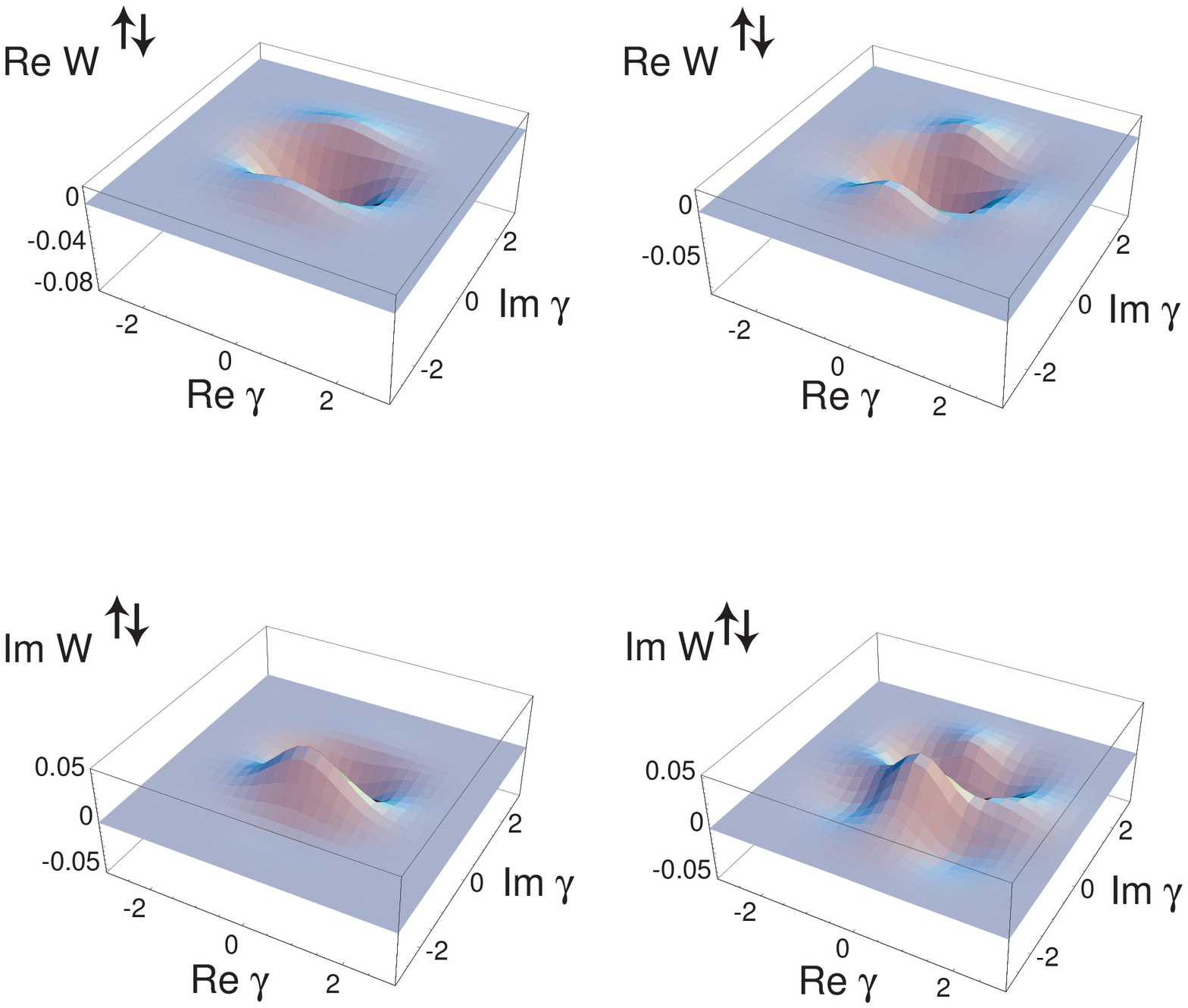,width=3.0in}}
\caption{\narrowtext 
Wigner function corresponding to the density matrix of
Fig.\ref{fig6} (true on the left and reconstructed on the right).
}
\label{fig8}
\end{figure}

Then, the Wigner functions corresponding to the density matrices
of Figs.\ref{fig4}, \ref{fig5} and \ref{fig6},
are shown in  Figs.\ref{fig7} and \ref{fig8}.

The reconstructed Wigner functions as well 
turn out to be quite faithful. 
We would
like to emphasize their particular shape.
In $W^{\uparrow\uparrow}$ there are two hills, one centered in 
${\rm Re}\,\gamma=-\alpha$ and the other in ${\rm Re}\,\gamma=\alpha$
coming from the random part of the state (\ref{eq:rhoal}),
instead, the pseudo singlet part of the state contributes
only to the bump in $-\alpha$, thus producing the asymmetric effect.
The opposite happens for $W^{\downarrow\downarrow}$.
The shape of $W_{\uparrow\downarrow}$ is due to the
quantum interference given by the entanglement between the two degrees
of freedom: in fact, in absence of entanglement 
$\rho^{\uparrow\downarrow}$ would just
be a replica of the diagonal parts 
$\rho^{\uparrow\uparrow}$
and $\rho^{\downarrow\downarrow}$.

\section{Physical Realization}

We now briefly discuss a system where the Werner-like
mixture could be synthesized and even measured.
It is a single electron trapped in a Penning trap~\cite{bg},
where two different degrees 
of freedom of the same particle can be
entangled and also measured.

An electron in a Penning trap is confined by the combination of a
homogeneous magnetic field along the positive $z$ axis and an
electrostatic quadrupole potential in the $xy$ plane~\cite{bg}.
The spatial part of the electronic wave function consists of three 
degrees
of freedom, neglecting the slow magnetron motion (whose 
characteristic
frequency lies in the kHz region), here we only consider the axial
and cyclotron motions, which are two harmonic oscillators radiating in 
the
MHz and GHz regions, respectively. On the other hand, the spin dynamics
results from the interaction between the magnetic moment of the 
electron
and the static magnetic field, so that the free Hamiltonian reads
as~\cite{bg}
\begin{equation}\label{eq:Hfree}
H_{\rm free}=\hbar\omega_z a_z^{\dag} a_z
+\hbar\omega_c a_c^{\dag} a_c
+\hbar\omega_s\sigma_3/2\,,
\end{equation}
where the indices $z$, $c$, and $s$ refer to the axial, cyclotron and 
spin
motions, respectively.

Here, in addition to the usual trapping fields, we consider an external
radiation field as a standing wave along the $z$ direction and 
rotating,
{\it i.e.} circularly polarized, in the $xy$ plane with frequency
$\Omega$~\cite{mmt}. To be more specific, we consider a standing wave 
within
the cylindrical cavity configuration~\cite{tan} with the 
(dimensionless)
wave vector $k$. Then, the interaction Hamiltonian~reads \cite{mmt}
\begin{eqnarray}\label{eq:Hnodip}
H_{\rm int}&=&
\hbar\epsilon\left[a_c e^{i(\Omega t+\varphi)}
+a_c^{\dag}e^{-i(\Omega t+\varphi)}\right]
\cos(k z +\chi)
\nonumber\\
&+&\hbar\zeta\left[
\sigma_-e^{i(\Omega t+\varphi)}
+\sigma_+e^{-i(\Omega t+\varphi)}
\right]
\sin(k z +\chi)\;,
\end{eqnarray}
where
$\sigma_{\pm}=(\sigma_1 \pm i \sigma_2)/2$, and
$z=a_{z}+a_{z}^{\dagger}$.
The phase $\chi$ defines the position of the center of the axial motion
with respect to the wave. Depending on its value the electron can be
positioned in any place between a node ($\chi=0$) and an antinode
($\chi=\pm\pi/2)$. 
Instead, the phase $\varphi$ is related to the initial direction
of the electric (magnetic) field in the $xy$ plane,
or to the phase of any reference field.
The two coupling constants $\epsilon$ and $\zeta$ 
are proportional to the amplitude of the applied radiation field.
Depending on $\Omega$ and $\chi$, 
the interaction Hamiltonian~(\ref{eq:Hnodip})
gives rise to different contributions at leading order in the Taylor 
expansion of $\sin(kz +\chi)$ and $\cos(kz +\chi)$.

We immediately recognize the possibility of implementing the  
transformations (\ref{eq:U}), (\ref{eq:D}) 
on the spin and cyclotron degrees of freedom by appropriately
exploiting the Hamiltonian (\ref{eq:Hnodip}).
For instance, $U$ can be realized by setting $\Omega=\omega_{s}$,
$\chi=0$, and then adjusting $\varphi$ and $\theta=\zeta t$.
Differently, $D$ can be realized by setting $\Omega=\omega_{c}$,
$\chi=-\pi/2$, and then adjusting $\beta=\epsilon t e^{-i\varphi}$.

These transformations easily allow to generate the disentangled 
components
of the mixture (\ref{eq:rhoal}), i.e., 
$|\uparrow\rangle_{1}|\pm\alpha\rangle_{2}$ and
$|\downarrow\rangle_{1}|\pm\alpha\rangle_{2}$,
starting from the typical initial state 
$|\uparrow\rangle_{1}|0\rangle_{2}$.

Instead, for what concerns the generation of the 
entangled fraction of the mixture 
(\ref{eq:rhoal}), we recall the procedure developed in 
Ref.\cite{michol}.
That is, we consider the possibility of introducing pulsed
standing waves through the microwave inlet~\cite{bg}
so that $\epsilon$, $\zeta$ become time dependent
and $\tilde{\epsilon}$, $\tilde{\zeta}$ indicate the pulse area
(the duration of the pulse is assumed to be much shorter than the
characteristic axial period, which is of the order of microseconds).
Then, nonclassical cyclotron states can be entangled with the spin 
states
through the following steps \cite{michol}. 

\begin{itemize}

\item{
First, we consider $\chi=0$, $\Omega=\omega_s$,
and a pulsed standing wave lasting $\Delta 
t_1=t_{1}-t_{0}=t_{1}$;}

\item{
Second, we allow a free evolution for a time $\Delta 
t_2=t_{2}-t_{1}=\pi/(2\omega_z)$;}

\item{
Third, we consider the action of another pulsed standing wave
with $\chi=-\pi/2$, $\Omega=\omega_c$, for a time 
$\Delta t_3=t_{3}-t_{2}$.} 

\end{itemize}

Finally, if we consider the initial axial state as 
a Gaussian state with momentum width much smaller than
${\tilde\zeta}$ (which is easily obtained in the case of
the ground state of the axial oscillator), we end up with
an evolution operator of the form
$D(\alpha\sigma_1)$,
where $\alpha$ is related to 
${\tilde\epsilon}$, ${\tilde\zeta}$ and $\varphi$.
It is then immediate to see that the initial state 
$|\uparrow\rangle_{1}|0\rangle_{2}$
may evolve with the aid of  
$D(\alpha\sigma_1)$ and a spin rotation into
\begin{equation}\label{eq:psesin}
\frac{1}{\sqrt{2}}
\left( |\downarrow\rangle_{1} |\alpha\rangle_{2}
-|\uparrow\rangle_{1} |-\alpha\rangle_{2} \right)\,.
\end{equation}
This state has been 
already discussed in Refs.~\cite{win1,haroche} 
and constitutes the
pseudo singlet component of
the mixture (\ref{eq:rhoal}).

Thus, at each run of the experiment the desired 
component of the Werner-like mixture can be synthesized,
thus allowing the generation of the state (\ref{eq:rhoal})
on average ensemble.

For what concerns the measurement,
the addition of a particular inhomogeneous magnetic field
(known as the magnetic bottle field \cite{bg})
to that already present in the trap, allows to perform
a simultaneous measurement of both the spin and the cyclotron 
excitations number. The useful interaction
Hamiltonian for the measurement process is
\begin{equation}\label{eq:Hbottle}
H_{\rm bottle}=\hbar \omega_{b}
\left[ a^{\dag}_{c}a_{c}+\frac{g}{2}\sigma_{3}\right] z^{2}\,,
\end{equation}
where the angular frequency $\omega_{b}$ is directly related 
to the strength of the magnetic bottle field.

Equation (\ref{eq:Hbottle}) describes the fact that the axial
angular frequency is affected both by the number of cyclotron
excitations $a_{c}^{\dag}a_{c}$ and by the eigenvalue of $\sigma_{3}$.
The modified (shifted) axial frequency can be experimentally measured
\cite{bg} after the application of the inhomogeneous magnetic
bottle field. One immediately sees that it assumes different values 
for every pair of eigenvalues of 
$a_{c}^{\dag}a_{c}$ and $\sigma_{3}$ due to the fact that 
the electron $g$ factor is slightly (but measurably \cite{bg})
different from $2$. 

However, prior such kind of measurement, one has to deal with
the transformations (\ref{eq:U}), (\ref{eq:D}) which can
be realized through the Hamiltonian (\ref{eq:Hnodip}) 
as discussed above.

Repeating this procedure many times 
allows us to recover the desired marginal distributions,
hence to sample the density matrix elements.

\section{Conclusions}

In conclusions, we have studied the properties
of a Werner-like mixture and a reliable method 
to achieve its tomographic characterization.
A useful system to investigate such states 
has been individuated in the trapped electron.
There are other candidate systems that offer the 
possibility to generate and manipulate the studied state.
We mention for example trapped ions \cite{nist}, 
or atoms in cavity quantum electrodynamics \cite{cqed}.
Moreover, in such systems the studied state might
involve two particles, or quite generally the two
subsystems could be spatially separated.

The experimental studies on this state might yield new insight
in the foundations of quantum mechanics and allow further progress in 
the 
field of quantum information.

\section*{Acknowledgements}
The authors warmly thank Dr. Mauro Fortunato
for helpful discussions.

\end{multicols}

\end{document}